\documentclass[prl,aps,twocolumn,superscriptaddress,balancelastpage]{revtex4-1}

\usepackage{latexsym,amssymb,bm,bbold,amsfonts,amstext,graphicx,bbm,bm,relsize,dsfont,times}

\usepackage{xcolor}
\usepackage{amsmath} 
\usepackage{amsthm} 
\usepackage{enumerate} 
\usepackage{float}

\usepackage[unicode=true, breaklinks=false, pdfborder={0 0 1}, backref=false, colorlinks=true, linkcolor=blue, citecolor=blue]{hyperref}

\newtheorem{definition}{Definition}

\newtheorem{theorem}[definition]{Theorem}
\newtheorem{corollary}[definition]{Corollary}

\definecolor{blue-violet}{HTML}{c13bff}

\usepackage[linesnumbered,ruled,vlined]{algorithm2e}
\usepackage{algcompatible,amsmath}
\algnewcommand\INPUT{\item[\textbf{Input:}]}%
\algnewcommand\OUTPUT{\item[\textbf{Output:}]}%

\newcommand{\tr}{\mathrm{Tr}}
\def\bra#1{\langle{#1}|}
\def\ket#1{|{#1}\rangle}
\def\braket#1{\langle{#1}\rangle}
\newcommand{\ketbra}[2]{\ket{#1}\!\bra{#2}}

\def\BraVert{\egroup\,\mid\,\bgroup}

\begin{document}

\title{Resource speed limits: Maximal rate of resource variation}
\author{Francesco Campaioli}
\email{francesco.campaioli@rmit.edu.au}
\affiliation{Chemical and Quantum Physics, and ARC Centre of Excellence in Exciton
Science, School of Science, RMIT University, Melbourne 3000, Australia}
\affiliation{School of Physics and Astronomy, Monash University, Victoria 3800, Australia}

\author{Chang-shui Yu}
\email{ycs@dlut.edu.cn}
\affiliation{School of Physics,
Dalian University of Technology,
Dalian, 116024, P. R. China}

\author{Felix A. Pollock}
\affiliation{School of Physics and Astronomy, Monash University, Victoria 3800, Australia}

\author{Kavan Modi}
\email{kavan.modi@monash.edu}
\affiliation{School of Physics and Astronomy, Monash University, Victoria 3800, Australia}
\affiliation{Institute for Quantum Science and Engineering, and Department of Physics, Southern University of Science and Technology, Shenzhen 518055, China}

\date{\today}

\begin{abstract}
Recent advances in quantum resource theories have been driven by the fact that many quantum information protocols make use of different facets of the same physical features, e.g. entanglement, coherence, etc. Resource theories formalise the role of these important physical features in a given protocol. One question that remains open until now is: \textit{How quickly can a resource be generated or degraded?} Using the toolkit of quantum speed limits we construct bounds on the minimum time required for a given resource to change by a fixed increment, which might be thought of as the power of said resource, i.e., rate of resource variation. We show that the derived bounds are tight by considering several examples. Finally, we discuss some applications of our results, which include bounds on thermodynamic power, generalised resource power, and estimating the coupling strength with the environment.
\end{abstract}

\maketitle

\makeatletter

\textbf{Introduction} --- Quantum information theory, over the past four decades and more, has unambiguously demonstrated that there are quantum information tasks without any classical counterparts. There are several physical quantum features responsible for such phenomena. For example, quantum entanglement is known to be a necessary resource for quantum communication protocols such as quantum teleportation~\cite{Bennett1993}, dense coding~\cite{Bennett1992}, and unconditional quantum encryption~\cite{Ekert1991}. It is also (highly likely to be) the key resource for quantum computing~\cite{arXiv:1907.08224}. Quantum discord~\cite{Ollivier2001, Henderson2001}, a type of nonclassical correlation, plays an important role in noisy quantum information processes, e.g. noisy quantum metrology~\cite{arXiv:1003.1174, arXiv:1504.02460, arXiv:1905.06101}. There many other important quantum features, such as coherence~\cite{Baumgratz2014}, magic states~\cite{magic}, and non-thermal states~\cite{Brandao2015}.

The multitude of quantum resources is not surprising. However, to organise the vast untapped resource fields, researchers have embarked on categorising quantum resources~\cite{Chitambar2019} using mathematical frameworks that are called \emph{quantum resource theories} \textbf{(QRTs)}. The core task of a QRT is to provide a quantitative understanding of a quantum feature, which is then used to identify the operations that generate, preserve, or degrade the resource, as well as the protocols that are required for its detection and effective application~\cite{Chitambar2019}. The success of this mathematical framework, which lies in its ability to reveal the common underlying structure of seemingly different resources, has sparked the rapid development of QRTs for a wide range of quantum features, such as asymmetry~\cite{Marvian2016}, 
coherence~\cite{Streltsov2017}, stabilizer and magic-state quantum computation~\cite{Veitch2014,Howard2017}, non-Gaussianity~\cite{Genoni2010}, continuous variable nonclassicality~\cite{PhysRevX.8.041038}, quantum measurements~\cite{guff2019}, quantum processes~\cite{berk-2019, wilde-2019-1, wilde-2019-2}, and generalised probability theories~\cite{PhysRevX.9.031053}.

Suppose we are given a quantum machine that runs on some quantum resource. Then an important operational problem is to quantify the rate of variation (production or degradation) of the resource in the quantum machine. For instance, Refs.~\cite{Uzdin2016, DelCampo2016} bound the rates of purity and coherence, respectively. More generally, is it possible to bound the rate of change in an arbitrary resource?  This is akin to computing bounds on the maximum power of a thermal machine. One approach is to bound the minimal time required to degrade or generate a fixed amount of resource. This can be done using a kind of time-energy uncertainty relation known as a \textit{quantum speed limit} \textbf{(QSL)}~\cite{Campaioli2020}; these have been used to study the limits of the rate of information transfer and processing~\cite{PhysRevResearch.2.013161}, charging and extraction power~\cite{Campaioli2017}, and other quantum information processing tasks~\cite{Giovannetti2011, Caneva2009, Carlini2006, Brody2015, Wang2015}, and have proven to be successful not only for applied quantum information~\cite{Murphy2010, Reich2012}, but also from a foundational standpoint~\cite{Deffner2017, Campaioli2018}.

In this Letter, we combine the framework of QRTs with the methods generally used for the derivation of QSLs to obtain two independent bounds on the minimal time required to vary a quantum resource, which we dub the \textit{resource speed limit} \textbf{(RSL)}. Our results are general, in that they make use of the quantum relative entropy \textbf{(QRE)} or Kullback-Leiber divergence measure, a universal measure for a large family of QRTs. We discuss the operational interpretation of our bounds and show how they naturally incorporate a penalty term in the form of the changes to the system's entropy. We then show how our bound can be used to obtain a traditional QSL and juxtapose its interpretation with that of our main results using quintessential resource theories, such as those of entanglement, coherence, and athermality. Within these examples, we show that the derived RSLs are tight, and can also outperform a QSL. We interpret such results in terms of the resource manifold and discuss their relevance and applicability.

\vspace{5pt}
\textbf{Relative entropy as resource measure} ---
For a given system with associated Hilbert space $\mathcal{H}$, a QRT is formally defined by a set of free states and free operations~\cite{Chitambar2019, Brandao2015a}. Naturally, the free states are those not owning the resource, i.e., those readily available; let their set be denoted by $\mathcal{F}\subset\mathcal{S}(\mathcal{H})$, with the latter the set of all quantum states. The resourceful states form the complement of $\mathcal{F}$; their set is usually denoted by $\mathcal{R} = \mathcal{S}(\mathcal{H})\setminus \mathcal{F}$. The set of free operations $\mathcal{O}$ is a unique collection of completely positive and trace-preserving (CPTP) operations on $\mathcal{S}(\mathcal{H})$ that cannot be used to increase a resource.

The task of quantifying a resource is accomplished by introducing a figure of merit to measure the \emph{value} of a state~\cite{Coecke2016, Gour2015, Chitambar2019}. A well defined resource measure $M:\mathcal{S}(\mathcal{H})\to\mathbb{R^+}$ is usually restricted by the following conditions: (i) $M$ vanishes for free states and is positive for resource states, i.e., $M(\sigma)=0 \leftrightarrow \sigma\in\mathcal{F}$ and $M(\rho)>0 \leftrightarrow \rho\in\mathcal{R}$; (ii) $M$ is a strong monotone, i.e., $\sum_{n} p_{n} M\left( \rho _{n}\right) \leq M\left( \rho \right) $ for any trace-decreasing free maps $\mathcal{K}_n: \rho \to \rho_n = \mathcal{K}_n(\rho) / p_n$ with $p_{n} = \mathrm{Tr}[\rho_n]$ and $\sum_n \mathcal{K}_n$ is trace preserving; and (iii) $M$ is convex, i.e., $\sum_{n}q_{n}M\left( \varrho _{n}\right) \geq M\left( \rho \right) $ for $\rho =\sum_{n}q_{n}\varrho _{n}$ with $q_n\geq 0$ and $\sum_n q_n =1$. The above conditions imply that the resource can neither increase under the action of free operations on average nor as a result of post-selection~\cite{Chitambar2019}, i.e., \emph{cherry-picking} of the outcomes of a measurement.

There are many different monotones to quantify the resource corresponding to a given a certain quantum feature. A particularly well-known monotone is the QRE~\cite{Vedral1998}, because it induces a well-defined resource measure independently of the chosen quantum feature:
\begin{gather}
\label{eq:re}
M(\rho) := \min_{\sigma \in \mathcal{F}}S\left( \rho \|\sigma
\right) = -S(\rho)-\tr[\rho\log\sigma^{\blacktriangledown}],
\end{gather}%
where $S(\rho )=-\tr[\rho \log {\rho }]$ is the von Neumann entropy, and where $\sigma^{\blacktriangledown}\in\mathcal{F}$ represents the free state that minimises the QRE with respect to the considered state $\rho$. For example, if $\mathcal{F}$ denotes the set of separable states, $M$ is the QRE of entanglement~\cite{Vedral2000}. Similarly, QRE can quantify other resources in non-classical states~\cite{Modi2011}, coherent states~\cite{Bu2017}, and non-Gaussian states~\cite{Genoni2008,Marian2013}. It can be easily checked that any quantum resource can be consistently characterised, subject to (i)--(iii), using QRE in this way~\cite{Vedral2018}. We therefore choose to work with QRE in this Letter. While our results are theory-independent, they can still be generalised to a larger class of metrics, e.g. $\alpha-$R\'enyi relative entropies~\cite{pmc}.

\vspace{5pt}
\textbf{Resource speed limit} --- We begin by considering the QRE as a resource monotone, represented by Eq.~\eqref{eq:re}. Let us denote the available set of initial states as $\mathcal{I} = \{\rho_0\}$. For generality, we let the system evolve according to the dynamics $\dot{\rho}_t = \mathcal{L}_t(\rho_t)$ prescribed by the quantum Liouvillian super-operator $\mathcal{L}_t$, which can describe both unitary evolution and dissipative dynamics (Markovian or non-Markovian)~\cite{Breuer2002}. The allowed dynamics map the initial set of states to a set of destination states 
\begin{gather}\label{liouville}
\mathcal{D} = \left\{\rho_\tau = \rho_0 + \int_0^\tau dt\: \mathcal{L}_t(\rho_t) : \rho_0\in \mathcal{I}, \tau \in \mathbb{R}^+\right\}.
\end{gather}
We will relate the dynamics to the change in the resource and the von Neumann entropy
\begin{align} \label{eq:MSdef}
  \Delta M &:= M(\rho_\tau)-M(\rho_0), \quad
  \Delta S := S(\rho_\tau)-S(\rho_0),
\end{align}
to present our first result.

\begin{theorem}
\label{th:penalty_bound_resource_variation}
Starting from a state $\rho_0\in \mathcal{I}$, the time $\tau$ required to arrive at a state $\rho_\tau \in \mathcal{D}$ with difference in resource value $\Delta M$, by means of the dynamics generated by the Liouvillian $\mathcal{L}_{t}$ is bounded as $\tau \geq T_{M}(\rho_0,\rho_\tau)$, with
\begin{align}
  \label{eq:penality_RSL}
  T_{M}(\rho_0,\rho_\tau) :=
  \frac{|\Delta M|}{\left<|-\tr[\mathcal{L}_t(\rho_t)\log\sigma^{\blacktriangledown}_x]-\dot{S}(\rho_t)|\right>_t},
\end{align}
where $x = \tau$ when $\Delta M\leq 0$ and $x = 0$ when $\Delta M\geq 0$.
\end{theorem}

\noindent
\textbf{Proof.} First we consider the case where $\Delta M\leq 0$. Substituting the inequality $S(\rho_0\|\sigma_\tau^{\blacktriangledown}) \geqslant S(\rho_0\|\sigma_0^{\blacktriangledown})$ into the expression for $-\Delta M$ we get
\begin{align}
-\Delta M &\leqslant 
  S(\rho_0\|\sigma_\tau^{\blacktriangledown})-S(\rho_\tau\|\sigma_\tau^{\blacktriangledown}), \\ 
  & = \Delta S + {\tiny \int_0^\tau} dt\ \tr\big[ \mathcal{L}_t(\rho_t)\log\sigma_\tau^{\blacktriangledown}\big].
\end{align}
The final line is obtained by using Eq.~\eqref{liouville}. We move $\Delta S = \int_0^\tau dt\: \dot{S}$ into the integral and take the absolute value of the integrand. Multiplying by $\tau/\tau$ and rearranging, we obtain bound~\eqref{eq:penality_RSL}.

For the $\Delta M \geq 0$ case, we take $S(\rho_\tau\|\sigma_0^{\blacktriangledown}) \geqslant S(\rho_\tau\|\sigma_\tau^{\blacktriangledown})$. The remainder of the proof follows similarly.\hfill $\blacksquare$

We now derive another RSL that is similar to the one in Th.~\ref{th:penalty_bound_resource_variation}. Here, the total entropy variation appears in the numerator rather than the denominator. We will later use this form to derive a QSL based on the QRE as a measure of distinguishability between states.

\begin{corollary}
\label{th:bound_resource_variation}
In the same settings as in Th.~\ref{th:penalty_bound_resource_variation}, the time $\tau$ is bounded as $\tau\geq \widetilde{T}_{M}$ with
\begin{gather}
  \label{eq:RSL}
  \widetilde{T}_{M}(\rho_0,\rho_\tau) :=
  \frac{|\Delta M+\Delta S|}{\left<|\tr[\mathcal{L}_t(\rho_t)\log\sigma^{\blacktriangledown}_x]|\right>_t},
\end{gather}
where $x = \tau$ when $\Delta M +\Delta S\leq 0$ and $x = 0$ when $\Delta M +\Delta S\geq 0$.
\end{corollary}
\noindent
The proof of this corollary follows as the proof of Th.~\ref{th:penalty_bound_resource_variation}, with the exception that $\Delta S$ is moved to the LHS of the inequality, instead of inside the integral. 

A few remarks are in order: First, we have heuristically observed that the bound~\eqref{eq:RSL} is often looser than bound~\eqref{eq:penality_RSL}. However, the two bounds coincide for unitary dynamics, and, in general, for nonunitary dynamics are independent~\footnote{Suppose the dynamics transform a pure product state to a mixed separable state. For a QRT of entanglement, bound~\eqref{eq:penality_RSL} is vanishing, while bound~\eqref{eq:RSL} is positive. Below we encounter several examples where the converse is true.}.

Second, for fixed $\rho_0$ and $\rho_\tau$, $\Delta M$ and $\Delta S$ are well defined, and one can use any QSL to bound $\tau$, including $\tau\geq T_M(\rho_0,\rho_\tau)$. However, the two RSLs above allow us to determine the absolute minimum time required to change a resource by some value $\Delta M = \mu$ by minimising over all pairs of initial states $\rho_0 \in \mathcal{I}$ and final states $\rho_\tau \in \mathcal{D}$
\begin{gather}
\label{eq:M_qsl}
T_{\mu} := \min_{\{\rho_0\in\mathcal{I}, \ \rho_\tau \in\mathcal{D}: \ \Delta M = \mu\}} T_{M}(\rho_0,\rho_\tau).
\end{gather}
As an example, this expression answers the question of how long it takes to generate $\mu$ ebits of entanglement. It is worth contrasting the above result with typical QSLs, where the numerator represents a notion of distinguishability (often using a distance measure) between the initial and the final state. In Th.~\ref{th:penalty_bound_resource_variation} and Cor.~\ref{th:bound_resource_variation}, the numerator quantifies a resource variation, while the denominator quantifies the rate of resource variation with respect to the nearest free state.

\begin{figure}[t]
\centering
\includegraphics[width=0.48\textwidth]{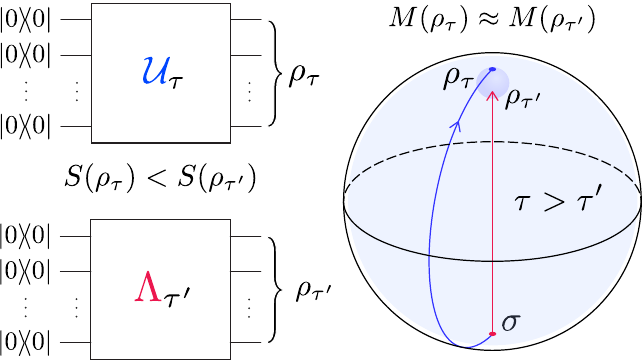}
\caption{\textbf{Entanglement and entropy generation} --- A global unitary evolution $U_t$ drives an initial free state $\sigma = \bigotimes_k \ketbra{0}{0}$ to a resourceful state $\rho_\tau = \mathcal{U}_\tau[\sigma]= U_\tau \sigma U_\tau^\dagger$, while keeping the entropy of the system $S(\rho_t)\equiv 0$ throughout the evolution. A non-unitary evolution $\Lambda_t$ takes $\sigma$ to another resourceful state $\rho_{\tau'}$ with resource $M(\rho_\tau)\approx M(\rho_{\tau'})$ comparable to that of $\rho_\tau$, but with higher entropy $S(\rho_\tau) < S(\rho_{\tau'})$. However, the latter evolution, in this depiction, is faster, i.e., $\tau>\tau'$ at the expense of increased entropy. Instantaneous and total entropy variations are accounted for by the bounds of Eq.~\eqref{eq:penality_RSL} and Eq.~\eqref{eq:RSL}. On the right, a Bloch-sphere  representation of the protocol for a 2-qubit system, with $\sigma=\ketbra{00}{00}$, and $\rho_\tau = \ketbra{\psi}{\psi}$, with $\ket{\psi} = (\ket{01}+\ket{10})/\sqrt{2}$.}
  \label{fig:entanglement_generation}
\end{figure}

Third, for a unitary process,  the term $v_x(t)=-\tr[\mathcal{L}_t(\rho_t)\log\sigma_x^\blacktriangledown]$, in both RSLs above, can be interpreted as the instantaneous speed of an evolution on an isentropic manifold of the state space. However, when entropy production cannot be avoided along the evolution, e.g. a non-unitary process, a penalty function, $\dot{S}(\rho_t)$, is subtracted from the speed $v_x(t)$ for bound~\eqref{eq:penality_RSL}. Analogously, for bound~\eqref{eq:RSL} the penalty function appears in the numerator as the change in the system's entropy. Moreover, in general, generating a resource will have an associated cost, which appears here as a change in entropy.  To interpret the penalty we direct the reader to Figure~\ref{fig:entanglement_generation}. Here, we aim to construct an entangled state using a quantum process. Our RSLs already show that this cannot be done instantaneously~\footnote{If you are thinking of \emph{getting lucky} with a measurement that, with some probability, might instantaneously collapse the system onto a resourceful state, think twice: Ideal projective measurement has an infinite energetic cost~\cite{Guryanova2020}, and weaker kinds of measurement cannot realise transformations faster than dictated by QSLs on average~\cite{GarciaPintosNJP}.}. In this example, we have a noisy computer that runs faster than a less noisy one. That is, to generate a fixed amount of entanglement, using program $\mathcal{U}_\tau$ will have a lower entropy cost than using program $\Lambda$. However, the run-time for $\mathcal{U}$ is longer than  that of $\Lambda_{\tau'}$, i.e., $\tau \ge \tau'$.

\vspace{5pt}
\textbf{Resource generation, degradation, and quantum speed limit} --- Now we consider two special cases that are of physical importance. First, we derive a bound on the time required to generate a resource. Next, we bound the time that is required to degrade the same resource. These cases correspond to experimental reality: the first case exemplifies the start of an experiment, which is initially in a fiducial state; the second case exemplifies its end, where the system will relax back to the fiducial state.

Let us consider a resource theory where there is only one free state $\sigma$. Moreover, let the initial state $\rho_0=\sigma$. We want to know how long it takes to reach a resourceful target state $\rho_\tau$. From Corollary~\ref{th:bound_resource_variation} we can obtain a bound on the time required to reach such resourceful state $\rho_\tau$, by imposing the condition $\sigma_0^\blacktriangledown = \sigma_\tau^\blacktriangledown = \sigma = \rho_0$ into Eq.~\eqref{eq:RSL}, to obtain
\begin{corollary}[Resource generation]
\label{cor:qsl_1}
The minimal time required to construct a resourceful state $\rho_\tau$, starting from the free state $\sigma = \rho_0$ is bounded from the below by
\begin{gather}
    \label{eq:res_gen}
    T_g (\sigma,\rho_\tau) :=  \frac{|S(\rho_\tau\|\sigma)+\Delta S|}{\left<|\tr[\mathcal{L}_t(\rho_t)\log\sigma]| \right>_t}.
\end{gather}
\end{corollary}

Within the same resource theory, suppose that instead we start from a resourceful state $\rho_0$, and we let our system evolve towards the free state $\sigma = \rho_\tau$. Like for Cor.~\ref{cor:qsl_1}, we can obtain a bound on the time required for the resource to degrade, by imposing the condition $\sigma_0^\blacktriangledown = \sigma_\tau^\blacktriangledown = \sigma = \rho_\tau$ into Eq.~\eqref{eq:RSL}, to obtain
\begin{corollary}[Resource degradation]
\label{cor:qsl_2}
The minimal time required to degrade a resource from state $\rho_0$ to the free state $\rho_\tau = \sigma$ is bounded from below by
\begin{gather}
    \label{eq:res_deg}
    T_d (\rho_0,\sigma) :=  \frac{|S(\rho_0\|\sigma)-\Delta S|}{\left<|\tr[\mathcal{L}_t(\rho_t)\log\sigma]| \right>_t}.
\end{gather}
\end{corollary}
\noindent
The proof of Cor.~\ref{cor:qsl_1} follows trivially substituting $\rho_0=\sigma_0^\blacktriangledown = \sigma_\tau^\blacktriangledown = \sigma$ into Eq.~\eqref{eq:RSL}. Similarly, the proof of Cor.~\ref{cor:qsl_2} follows trivially substituting $\rho_\tau=\sigma_0^\blacktriangledown = \sigma_\tau^\blacktriangledown = \sigma$ into Eq.~\eqref{eq:RSL}.

Let us notice that, for some free states, the degradation (generation) process can take an infinitely long time. For example, while entanglement can vanish suddenly, i.e., in a finite time, other resources such as discord, coherence, and athermality may not vanish in finite time. In this case, the bounds of Eqs.~\eqref{eq:res_gen} and~\eqref{eq:res_deg} can end up being loose. To circumvent this problem, one can select a final state $\rho_\tau\neq\sigma$ such that $M(\rho_\tau)>M(\sigma)$, or an initial state $\rho_0\neq\sigma$ such that $M(\rho_0)>M(\sigma)$, for Eqs.~\eqref{eq:res_gen} and~\eqref{eq:res_deg}, respectively. Now, the numerators of Eqs.~\eqref{eq:res_gen} and~\eqref{eq:res_deg} change by a \textit{small} quantity for \textit{small} deviations from the free state $\sigma$. 

The above two corollaries imply a QSL for the evolution between any two states $\rho_0$ and $\rho_\tau$ by interpreting the states as the unique free state of two resource theories. The QSL can be taken as the maximum of the two bounds of the corollaries above, $T_g(\rho_0,\rho_\tau)$ and $T_d(\rho_0,\rho_\tau)$. 
While the QRE is not a distance~\footnote{The QRE does not respect symmetry or the triangle inequality.}, it is a valid a measure of distinguishability between two quantum states~\cite{Vedral2002}. We can take advantage of the asymmetry of the QRE to express the QSL using
\begin{gather}
    \label{eq:QSL_relative_entropy}
    T(\rho_0,\rho_\tau):=\frac{|S(\rho_0||\rho_\tau)-\Delta S|}{\langle |\tr[\mathcal{L}_t(\rho_t)\log\rho_\tau]|\rangle_t}.
\end{gather}

\begin{corollary}[Quantum speed limit]
\label{cor:qsl_3}
The time $\tau$ required to evolve between any two states $\rho_0$ and $\rho_\tau = \sigma$ by means of the dynamics generated by the Liouvillian $\mathcal{L}_t$ is bounded as
\begin{gather}
    \label{eq:RL_QSL}
    \tau\geq\max\{T(\rho_0,\rho_\tau),T(\rho_\tau,\rho_0)\}.
\end{gather}
\end{corollary}
\noindent
In the next section, we consider some important examples of resource degradation dynamics to calculate the bounds~\eqref{eq:penality_RSL},~\eqref{eq:RSL}, and~\eqref{eq:RL_QSL}, and discuss their tightness, attainability, and interpretation.

\vspace{5pt}
\textbf{Examples: Entanglement, discord, and coherence} --- 
For our first example, we study entanglement degradation working within the QRT of entanglement. To this end, we compute (analytically and numerically) the two RSL bounds $T_M$ in~\eqref{eq:penality_RSL} and $\widetilde{T}_M$ in~\eqref{eq:RSL}, as well as the QSL $T$ in~\eqref{eq:RL_QSL}. For reference, we compare these bounds to the tight QSL $T_D$ introduced in Ref.~\cite{Campaioli2019}, which has been shown to outperform other QSLs for open quantum evolution. Each bound is compared with the evolution time $\tau$. Note that it is only meaningful to compare RSLs and QSLs for a given process.

To be specific, we consider a two-qubit system initialised in the Werner state
\begin{gather}
  \label{eq:entangled_plus}
  \rho_W(p):=\frac{p}{4}\mathbb{1}+(1-p)\ketbra{\phi^+}{\phi^+},
\end{gather}
where $\ket{\phi^+}=(\ket{00}+\ket{11})/\sqrt{2}$. The separable state $\sigma_W(p)$ that minimises the QRE with respect to $\rho_W(p)$ is obtained by dephasing the above state in the computational basis~\cite{Vedral1997}. The closest separable state to a Werner state is also the closest classically correlated state and the closest incoherent state. Hence, our example automatically includes the cases of QRE of discord and coherence.

\textit{Dephasing.} To model resource degradation, we first consider the pure dephasing channel with action $\rho_t = \Lambda_t^{\nu}[\rho_0]$ parameterised by $\nu(t) = \exp[-\gamma t]$, where $\gamma$ is the phase-relaxation rate~\cite{Yu2003}. The action of this channel on the initial state being the Werner state can be simply described by the decay of the off-diagonal terms $\braket{11| \rho_t |00} = \braket{00| \rho_t |11} = \nu(t)(1-p)$. As this channel leads to the most direct resource degradation, we expect our bounds to reveal the optimality of such resource variation dynamics. Indeed, upon analytically calculating the aforementioned bounds we obtain $T_M= \widetilde{T}_M=T=T_D = \tau$. These results confirm that these bounds are tight and attainable~\footnote{In this case {$\widetilde{T}_M$} is calculated in the limit of {$\epsilon\to0$} for the \emph{almost}-free state {$\sigma_W(p)+\epsilon(\ketbra{00}{11}+\ketbra{11}{00})$}.}.

\begin{figure}
\centering
\includegraphics[width=0.5\textwidth]{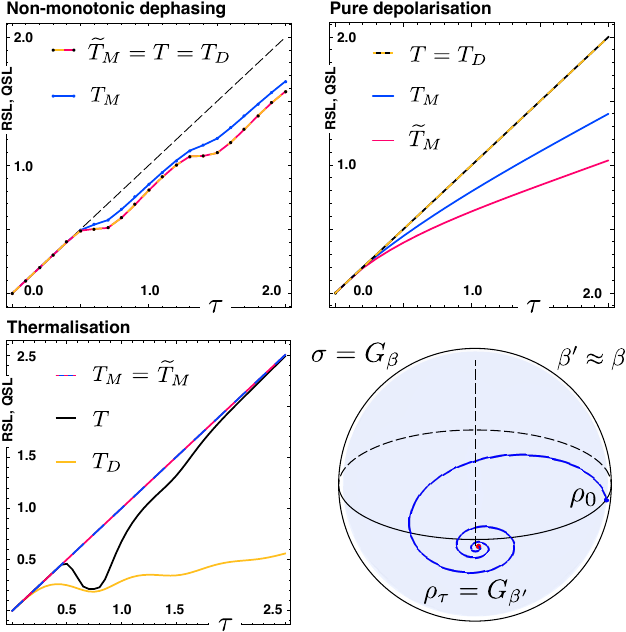}
\caption{\textbf{Examples} --- Numerical evaluation of RSL and QSL bounds for (\textit{top left}) non-monotonic dephasing with $\gamma = 0.2$, $k=4$, $p=0.5$ (hierarchy holds for any value of the parameters, $T_M$ depends on the mixedness parameter $p$, while $T$ and $\widetilde{T}_M$ are insensitive to it); (\textit{top right}) pure depolarisation with $\gamma = 1$, $p=0.9$ (hierarchy holds for any value of the parameters); (\textit{bottom left and right}) thermalisation with $\rho_0 = (\mathbb{1}+\sigma_y)/{2}$, $\omega = 4$, $\gamma=2$, $\beta=1/5$ (hierarchy depends on the parameters).}
\label{fig:bounds}
\end{figure}

\textit{Depolarisation.} We next consider the pure depolarisation channel to model entanglement degradation. This channel maps Werner states onto Werner states and can be simply defined in terms of a time-dependent mixedness parameter $p(t) = 1 - \exp[-\gamma t+\log(1-p_0)]$, where $\gamma$ is the depolarisation rate. The QSL between pairs $\rho_W(p_0)$ and $\rho_W(p_\tau)$ can be analytically shown to be $T=T_D=\tau$, i.e., both QSLs are tight. We numerically computer the RSL bounds and find $\widetilde{T}_M<T_M<T=T_D=\tau$ (for all values of $p_0$ and $\gamma$), as shown in Fig.~\ref{fig:bounds} (\textit{top right}). As opposed to the case of pure dephasing, the lack of tightness for the RSLs indicates the pure depolarisation channel is not the most direct way of degrading entanglement. This is in spite of the fact that this process does naturally correspond to the fastest evolution between a pair of $\rho_W$ states, as indicated by the tightness of the QSLs.

\textit{Non-Markovian processes.} For both channels considered above, the decay rate is constant and the decay is monotonic. We now relax this condition and reconsider the above two channels with non-monotonic decay rates, i.e., non-Markovian processes. The lack of a clear connection between QSLs and non-Markovianity was recently argued in Ref.~\cite{arXiv:1907.05923}. This agrees with Ref.~\cite{Campaioli2019}, which showed that the path between two states can be shorter or longer for non-Markovian processes when compared to the shortest Markov process.

For the case of non-monotonic dephasing, for which $\nu(t)=\exp[-\gamma(t+\sin^2(kt)/k)]$ with $k>\gamma$, we numerically calculate the bounds to obtain  $ \widetilde{T}_M = T = T_D < T_M < \tau$ (for non-trivial choices of parameters $\gamma$ and $k$), as shown in Fig.~\ref{fig:bounds} (\textit{top left}). These results confirm that the bounds $T_M$ and $\widetilde{T}_M$ are able to single out sub-optimal resource variation orbits. Finally, for the case of non-monotonic depolarisation, where $p(t)=(1-\exp[-\gamma (t+\sin^2(kt)/k)+\log(1-p_0)])$ with $k>\gamma$, we calculate all the bounds numerically to obtain $\widetilde{T}_M<T_M<T=T_D<\tau$ (for all non-trivial choices of parameters $p_0$, $\gamma$, and $k>\gamma$).

\vspace{5pt}
\textbf{Example: Thermal states} --- We now briefly look at the resource theory of athermality, for which the free states correspond to thermal states, i.e.,  the Gibbs canonical ensemble. For simplicity, we consider a single two-level system with internal Hamiltonian $H_0=\omega\,\sigma_z$, that evolves under the effect of a large heat bath at inverse temperature $\beta$. The dynamics of the system is governed by the Lindblad master equation
\begin{gather}
    \label{eq:linblad_thermal}
\dot{\rho}_t = - i [H_0,\rho_t]
+\!\!\!\!\!\!\sum_{j=\{+,-\}}\!\!\! \Gamma_{j} \left(2\sigma_{j} \rho_t \sigma_{j}^\dagger + \{\sigma_{j}^\dagger\sigma_{j},\rho_t\} \right),
\end{gather}
where $\sigma_{\pm} = (\sigma_x \pm i \sigma_y)/2$, $\Gamma_{+}= \gamma\frac{N}{2}$ and $\Gamma_{-}= \gamma\frac{N+1}{2}$ with $N=(\exp\{2\:\omega/\beta\}-1)^{-1}$. Here, the rate $\gamma$ is analogue to the rate of spontaneous emission for an atom-cavity interaction~\cite{Breuer2002}. The dynamical map $\rho_t = \Lambda_t^\beta[\rho_0]$ obtained from Eq.~\eqref{eq:linblad_thermal} asymptotically maps any initial state $\rho_0$ to the thermal state $G_\beta = \exp\{-\beta H_0\}/\mathcal{Z}$, with $\mathcal{Z}=\tr[\exp\{-\beta H_0\}]$. 

As before, we calculate bounds numerically to obtain $T_D \leq T\leq T_M= \widetilde{T}_M=\tau$, where the equal sign for the first inequality holds when $[\rho,H_0]=0$, and the strict inequality holds for any initial state that does not commute with the internal Hamiltonian, even though $T$ and $\tau$ can be arbitrarily close for the right choice of $\rho_0$. These results are of straightforward interpretation: The dynamics described by $\Lambda_t^\beta$ monotonically decreases the athermality of any initial state in the most direct way (due to the tightness of RSLs $T_M$ and $\widetilde{T}_M$) but does not connect $\rho_0$ to $\rho_\tau$ in the most direct way (due to the looseness of QSL $T$, as well as $T_D$ of Ref.~\cite{Campaioli2019}). We have depicted this in Fig.~\ref{fig:bounds} (\textit{bottom left and right}).

\vspace{5pt}
\textbf{Conclusions.} ---
The above examples clarify the role of the RSL bounds derived in this Letter, and contrast them with traditional QSls, such  as that in Ref.~\cite{Campaioli2019} and the one derived here. Our examples show that the RSLs are tight and attainable when the system traverses on an orbit that varies the resource in the most direct way. For the pure dephasing example, where the orbit is time-optimal, the RSLs coincide with QSLs. In contrast, pure depolarisation exemplifies when the QSLs outperform the RSLs. Namely, when the evolution between two states is optimal but the variation of the resource is sub-optimal. This example highlights that a pure depolarisation channel is not the fastest way of degrading entanglement, discord, or coherence. These examples may suggest that QSLs are generally tighter than RSLs, However, our third example shows otherwise. For thermalisation, the RSLs reveal that every orbit generated by the dynamics given in Eq.~\eqref{eq:linblad_thermal} is optimal for degrading athermality. In this case, the RSLs outperform traditional QSLs when the initial state is not already thermal. This is because QSLs evaluate the minimal time required to cover the geodesic connecting the initial and final states, rather than the variation $\Delta M$ of the considered resource, which follows the spiral path shown in Fig.~\ref{fig:bounds}.

The differing roles and performance of RSLs and QSLs can be traced back to their the construction. QSLs are typically formulated upon a notion of distance on the space of states, which operationally corresponds to a measure of distinguishability~\cite{Pires2016, Deffner2017, Campaioli2019}. While they reveal the \emph{optimality} of an evolution between two quantum states with respect to the considered metric~\cite{Deffner2017, Campaioli2019}, they do not necessarily provide the minimal time required for varying a resource under some quantum dynamical process. In contrast, saturating the RSL indicates that the underlying process is optimal at varying said resource. And, in such instances, an RSL will yield a better estimate for minimal time than any QSL.

There are several distinct directions in which the studies of RSLs can be extended. The rate of variation of the resource can be thought of as the resource power. By the same logic we may think of resource generation and degradation in Corrs.~\ref{cor:qsl_1} and~\ref{cor:qsl_2} as `resource work' and `resource heat'. Such constructions pave the path for defining efficiency in using or creating a resource \textit{\`a la} thermodynamics. Our methods are easily extendable, so that the RSLs can be generalised to the full class of $\alpha$-R\'enyi relative entropies~\cite{pmc}, which form a family of second laws of thermodynamics~\cite{Brandao2015}. Another research avenue could involve designing analytical and numerical methods to look for fast and efficient resource variation protocols, similarly to the approaches in Refs.~\cite{Wang2015, Campaioli2019b} for the case of unitary evolution. The RSLs may also help to bound the coupling strength with the environment by estimating the rate of degradation for some resources (e.g. entanglement or coherence). Finally, there are several classical resource theories, which in conjunction with classical speed limits~\cite{ShanahanPRL2018, OkuyamaPRL2018, ShiraishiPRL2018}, can be used to develop classical RSLs.

\begin{acknowledgments}
\textbf{Acknowledgments.}--- FC would like to acknowledge the Postgraduate Publication Award for partially funding this research. This research was also funded in part by the Australian Research Council under grant number CE170100026. CSY is supported by the National Natural Science Foundation of China, under Grant No.11775040. KM is supported through Australian Research Council Future Fellowship FT160100073. K.M. acknowledges support from the Australian Academy of Technology and Engineering via the 2018 Australia China Young Scientists Exchange Program and the 2019 Next Step Initiative. KM thanks {\L}ukasz Rudnicki for an insightful discussion.
\end{acknowledgments}

\bibliography{main}
\end{document}